\documentclass[a4paper,11pt]{article}
\usepackage{pos}
\usepackage{multirow}

\newcommand{\csw}{c_\mathrm{SW}}

\def\gsim{\mathrel{\raise2pt\hbox to 8pt{\raise -5pt\hbox{$\sim$}\hss{$>$}}}}
\def\rsim{\mathrel{\raise2pt\hbox to 8pt{\raise -5pt\hbox{$\sim$}\hss{$>$}}}}
\def\lsim{\mathrel{\raise2pt\hbox to 8pt{\raise -5pt\hbox{$\sim$}\hss{$<$}}}}

\newcommand{\beq}{\begin{equation}}   
\newcommand{\eeq}{\end{equation}}   
\newcommand{\beqn}{\begin{eqnarray}}   
\newcommand{\eeqn}{\end{eqnarray}}

\title{$K$- and $D_{(s)}$-meson leptonic decay constants with physical light, strange and charm quarks by ETMC}

\ShortTitle{$K$- and $D_{(s)}$-meson leptonic decay constants by ETMC}

\author*[a]{P. Dimopoulos} 
\author[b]{R. Frezzotti}
\author[c]{M. Garofalo}
\author[d]{S. Simula}

\affiliation[a]{Dipartimento di Scienze Matematiche, Fisiche e Informatiche, Universit\`a di Parma, Parma, Italy}
\affiliation[b]{Dipartimento di Fisica and INFN, Università di Roma “Tor Vergata", I-00133 Rome, Italy}
\affiliation[c]{HISKP (Theory), Rheinische Friedrich-Wilhelms-Universität Bonn, Germany} 
\affiliation[d]{Istituto Nazionale di Fisica Nucleare, Sezione di Roma Tre, Italy}

\emailAdd{petros.dimopoulosATunipr.it}

\abstract{We present a lattice QCD computation and preliminary results for the leptonic decay constants of the pseudoscalar mesons $K$, $D$ and $D_{s}$ in the isosymmetric QCD limit. The computation is based on simulations of $N_f=  2+1+1$  dynamical quarks performed  by the Extended Twisted Mass Collaboration (ETMC), where the light, strange and charm quark masses are all tuned at their physical values. We also present preliminary  unitarity checks for the first and second rows of the Cabibbo-Kobayashi-Maskawa matrix.}

\FullConference{%
 The 38th International Symposium on Lattice Field Theory, LATTICE2021
  26th-30th July, 2021
  Zoom/Gather@Massachusetts Institute of Technology
}

\begin{document}
\maketitle

\section{Introduction}
Of the most important  hadronic inputs for obtaining estimates of the  Cabibbo-Kobayashi-Maskawa (CKM) matrix elements (ME) are the values of the leptonic decay constants of pseudoscalar (PS) mesons. In these proceedings  we present a  high precision lattice QCD (LQCD) calculation  for the $K$, $D$ and $D_s$ PS-meson decay constants using $N_f =  2+1+1$ gauge ensembles generated by  the Extended Twisted Mass  Collaboration (ETMC). 

\section{Lattice action and simulations}

We employ the twisted mass (tm) fermionic formulation that ensures automatic O$(a)$-improvement  for all observables  as far as tuning at maximal twist is in place~Refs~\cite{FrezzoRoss1, Frezzotti:2003xj}. Moreover as it has been shown in Refs~\cite{Alexandrou:2018egz, ETM:2015ned,  Dimopoulos:2009es, Becirevic:2006ii} the inclusion of the clover term in the maximally twisted fermionic action provides  the beneficial property of reduced O$(a^2)$ cutoff and isospin breaking effects. 

ETMC has performed $N_f= 2+1+1$ dynamical quark simulations employing the sea quark action written as $S_{sea}  = S_{g}^{\text{Iwa}} + S_{tm}^\ell  + S_{tm}^h$, 
where $S_{g}^{\text{Iwa}}$ is the Iwasaki improved gauge action~\cite{Iwasaki:1985we} and 
\begin{align}   
     S_{tm}^\ell &= \sum_x \bar{\chi}_\ell(x)\left[ D_W(U) + \frac{i}{4} \csw \sigma^{\mu\nu}
    \mathcal{F}^{\mu\nu}(U) + m_{0\ell} + i \mu_\ell \tau^3 \gamma^5 
  \right] \chi_\ell(x)\, , \label{Eq:tmlight}  \\ 
 S_{tm}^h &= \sum_x \bar{\chi}_h(x)\left[ D_W(U) + \frac{i}{4} \csw \sigma^{\mu\nu}
    \mathcal{F}^{\mu\nu}(U) + m_{0h} - \mu_\delta \tau_1 + i \mu_\sigma \tau^3 \gamma^5
  \right] \chi_h(x)\, ,  \label{Eq:tmheavy}
     \end{align}
are, respectively,  the fermionic actions used in the light ($S_{tm}^\ell$) and strange-charm ($S_{tm}^h $) quark sectors. Notice that in Eq.~(\ref{Eq:tmlight})  $\chi_{\ell} = (u, d)^T$ represents the light quark doublet, while the degenerate light twisted and the (untwisted) Wilson quark masses are denoted by $\mu_{\ell}$ and $m_{0\ell}$, respectively. For the heavy quark action, Eq.~(\ref{Eq:tmheavy}),  the doublet $\chi_h = (s, c)^T$ represents  the mass non-degenerate strange and charm quarks. Here $m_{0h}$ denotes the (untwisted) Wilson quark mass while the parameters $\mu_{\delta}$ and $\mu_{\sigma}$ in combination with the presence of the Pauli matrices $\tau_1$ and $\tau_3$ lead to quark mass non-degeneracy, see Ref.~\cite{Frezzotti:2003xj}. In both equations  $D_W(U)$ is the massless Wilson-Dirac operator, and the Sheikoleslami-Wohlert improvement term, $\csw \sigma^{\mu\nu} \mathcal{F}^{\mu\nu}(U)$,  has been included for the reason already explained above. The value for the clover parameter $\csw$ is set by using the 1–loop tadpole boosted estimate as presented in Ref.~\cite{Aoki:1998qd}. The condition of maximal twist is achieved by tuning the hopping parameter $\kappa$ for the untwisted Wilson quark mass such as  
$m_{0\ell} = m_{0h} = m_{crit}$. Details about the lattice action and the algorithmic setup are presented in Refs~\cite{Alexandrou:2018egz, Alexandrou:2021bfr, JacobLAT21}. 

Simulations have been carried out reaching the physical mass values of both light and heavy 
(strange and charm) quarks. As for the latter the sea quark mass parameters ($\mu_{\sigma}$ and $\mu_{\delta}$) have been tuned so that the two phenomenological conditions $m_c/m_s = 11.8$ and $m_{D_{s}}/f_{D_{s}} = 7.9$~\cite{FlavourLatticeAveragingGroup:2019iem} are accurately reproduced by each of the  $N_f= 2+1+1$ ensembles.

In order to avoid O$(a^2)$ mixing effects in the physical observables involving the heavy quarks (strange and charm),  owing to the form of the sea quark action of Eq.~(\ref{Eq:tmheavy}), we opted for a non-unitary lattice setup. Therefore in the valence sector we employ the Osterwalder-Seiler fermionic regularisation~\cite{Osterwalder:1977pc} which treats the strange and charm quarks in a flavour diagonal way.  The valence action in the strange and charm sectors is given by: 
\begin{equation}
 S_{val}^{f} =   \sum_{x} \bar{\chi}_{f}^{val}(x) \left(D_{W}^{cr}(U) + \frac{i}{4} \csw \sigma^{\mu\nu}
    \mathcal{F}^{\mu\nu}(U)  + i \gamma_5 \mu _f \right) \chi_{f}^{val}(x)\,\, , f = s,c, 
\end{equation}
where $D_{W}^{cr}(U) \equiv D_W(U)|_{m_0 = m_{crit}}$ is the critical Wilson-Dirac operator and $\chi_{f}^{val}$ denotes single quark flavour field. It has been shown in Ref.~\cite{Frezzotti:2004wz} that this kind of mixed action preserves the automatic O$(a)$-improvement of physical observables, i.e. lattice artifacts, including those violating unitarity, scale as O$(a^2)$ implying that unitarity is safely recovered in the continuum limit. 

In this study we use $N_f = 2+1+1$ simulations performed at three values of the lattice spacing in the range $[0.69, 0.95]$ fm and at several pseudoscalar mass values spanning from the physical pion mass up to 350 MeV. In Table~\ref{Tab:simdetails} essential simulation details are presented. For the computation of $w_0/a$ and the gradient-flow $w_0$-determination, we refer the reader to  Refs~\cite{Alexandrou:2021gqw,  Alexandrou:2021bfr,    
 BartekLAT21}. The scale setting is performed using the isosymmetric QCD value $f_{\pi}^{isoQCD} = 130.4(2)$~\cite{Aoki:2016frl}. 
 
 The present lattice computation is performed in the isoymmetric QCD limit. However future work based on the same gauge ensembles is planned in order  to take into account isospin breaking effects along the lines of the work of  {\it e.g.}  Ref~\cite{DiCarlo:2019thl}. 

\begin{table}[h!]
		\begin{center}
		\scalebox{0.85}{
		\begin{tabular}{|c|c|c|c|c|c|}
			\hline
			$\beta$ & Ens. & $(L,T)$  & $M_{\pi}$ (MeV) & \# meas. & $w_0/a$ \\
			\hline
			\multirow{4}{*}{1.726}& cA211.12.48 & (48,96) & 167 & 322 & \multirow{4}{*}{1.8355(35)}  \\
			 & cA211.30.32 & (32,64) & 261 & 1237 & \\
			 & cA211.40.24 & (24,48) & 302 & 662 & \\
			  & cA211.53.24 & (24,48) & 346 & 628 & \\
			\hline
			
			\multirow{4}{*}{1.778}& cB211.072.64 & (64,128) & 137 & 374 & \multirow{4}{*}{2.1300(16)} \\
			& cB211.14.64 & (64,128) & 190 & 437 & \\ 
		     & cB211.25.48 & (48,96) & 253 & 314 &  \\
		       & cB211.25.32 & (32,64) & 253 & 400 &  \\
		    
			\hline
			
			{1.836} & cC211.06.80 & (80,160) & 	134 & 401 &     2.5045(17)  \\
			        & cC211.20.48  & (48,96)  & 246  & 890 & \\

			\hline
		\end{tabular}
	}
	\end{center}
	\caption{Simulation details for the $N_f=2+1+1$ ensembles by ETMC.} \label{Tab:simdetails}
	\end{table}

\section{Determination of pseudoscalar meson decay constants with tmQCD}
In the maximal tm (Mtm) formulation of LQCD the computation of pseudoscalar decay constants does not require any  (re)normalisation constant thanks to the existence of a conserved current~\cite{FrezzoRoss1, Frezzotti:2000nk}. It is thus sufficient to employ correlation functions of the type
$C_{PP}(t) = (1/L^3) \sum_{\vec{x}, \vec{y}} \langle 0 | P_{ff'}(\vec{x}, t) P^{\dagger}_{ff'}(\vec{y}, 0) | 0 \rangle ,$ 
where   $P_{ff'}(x) = \bar{\chi}_{f}^{val}(x) \gamma_5 \chi_{f'}^{val}(x)$ (with flavours $\{f,f'\} = \{\ell, s, c \}$) is the pseudoscalar density operator.
Then the leptonic decay constant of a PS-meson with mass denoted by $M_{ps(ff')}$ made out of valence quark flavours with bare masses $\mu_{f}$ and $\mu_{f'}$  is given by
\begin{equation}  
  f_{ps} = (\mu_f + \mu_{f'}) \dfrac{\langle 0 | P_{ff'} | ps \rangle}{M_{ps(ff')}  \sinh(M_{ps(ff')})}~,  
\end{equation}
which is automatically O$(a)$-improved owing to the maximal twist condition ($m_0^{val} = m_{0\ell} = m_{0h} = m_{crit}$).

In the present computation we make use of the quark mass results for the $u/d$, $s$ and $c$  presented in Section V of Ref.~\cite{Alexandrou:2021gqw}   and corresponding to the meson sector analysis. 
In this way we can employ the same mesonic correlation functions as in Ref.~\cite{Alexandrou:2021gqw} and
hence determine correctly the error propagation owing to the quark masses’ uncertainties.
We recall that in the framework of Mtm LQCD the renormalised quark mass of a quark
flavour $f$ is given by $m_f = \mu_f / Z_P$, 
where $Z_P$ is the  renormalization constant for the pseudoscalar density operator, the determination of which has been presented in Refs~\cite{Alexandrou:2021gqw, MatteoLAT21}.

For the statistical and fit error analysis we have employed the jackknife method. We have determined  the continuum limit values at the physical light quark mass $u/d$ by making use of simultaneous continuum and chiral fits.  
For the estimation of the various sources of systematic uncertainty we repeat our analysis by employing different kinds of fit ans\"atze regarding the  chiral extrapolation/interpolation to the physical light quark mass $u/d$. Furthermore we perform several analyses by using  data combinations corresponding to two out of three lattice spacings and also  by employing different determinations for $Z_P$ that differ by O$(a^2)$ effects.  
For a given decay constant we thus obtain a distribution of results, with each result corresponding to a different analysis and a
total number of analyses ranging from 32 to 96 (depending on the considered decay constant). 
From such a distribution the mean value and the uncertainty of the final result are estimated
using the combination method and formulae discussed in Sec. V of Ref.~\cite{Alexandrou:2021gqw} (see there Eqs (38)--(43)).

\section{Determinations of $f_K$ and $f_K/f_{\pi}$}

In the determination of $f_K$  we first interpolate the decay constant estimates to the strange quark mass and then we employ simultaneous continuum and chiral fits of the quantity $f_{s\ell}$ against the light quark mass $m_{\ell}$. We make use of two fit ans\"atze, namely the  next-to-leading order (NLO) SU(2) ChPT formula,  $f_{s \ell}  = P_{0} \, ( 1 - (3/4)\xi_{\ell} \log \xi_{\ell} + P_{1} \xi_{\ell} + 
P_{2} a^2 ) \,  \text{K}_{f_{K}}^{\text{FSE}}$ and a polynomial quadratic fit of the form $f_{s\ell}  = Q_{0}' \, \left( 1 + Q_{1}' m_{\ell} + Q_{2}' m_{\ell}^2 + Q_{3}' a^2 \right)
\,  \text{K}_{f_{K}}^{\text{FSE}}$. In the first one it is set $\xi_{\ell} = (2B_0 m_{\ell}) / (4 \pi f_0)^2$ where $B_0$ and $f$ are the SU(2) ChPT low-energy constants (LECs)  obtained from the quark mass analysis, see Ref~\cite{Alexandrou:2021gqw}. The factor $\text{K}_{f_{K}}^{\text{FSE}}$ represents the estimation for the (small) correction to our data due to finite size volume effects (FSE)  following~\cite{Colangelo:2005gd}.

\begin{figure*}[h!]
\centering
\begin{tabular}{lcr}
    \includegraphics[width=0.47\linewidth]{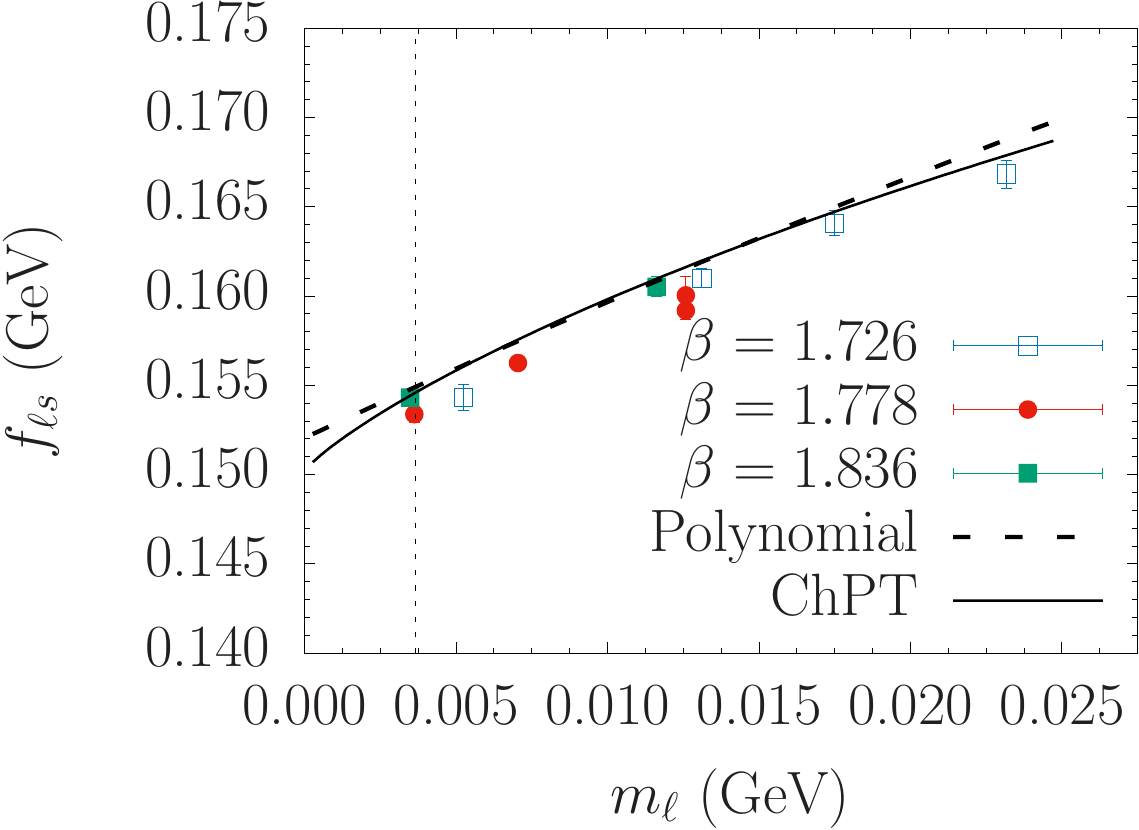}& &
    \includegraphics[width=0.46\linewidth]{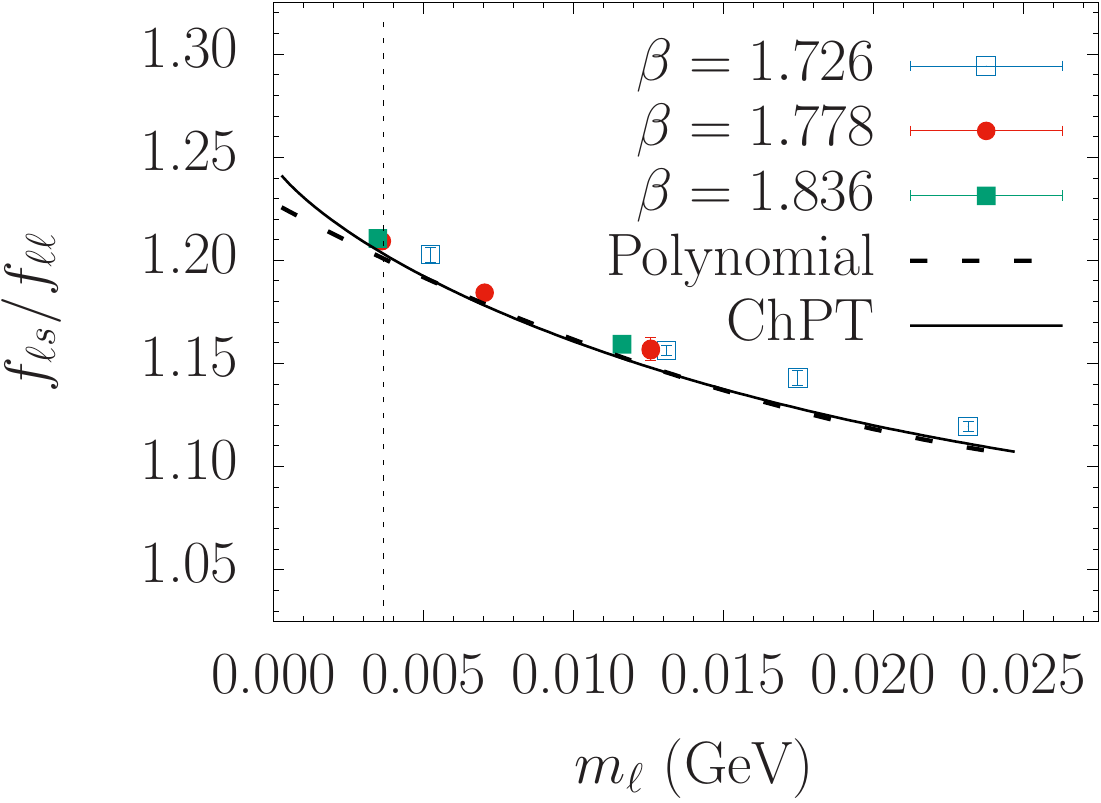}      \\
    \end{tabular}
\caption{Plots showing  simultaneous chiral and continuum fits (continuum limit curves are shown) for $f_K$ (left) and $f_K/f_{\pi}$ (right) against the renormalised light quark mass $m_{\ell}(\overline{\text{MS}}, 2~\text{GeV})$. Ans\"atze for chiral and polynomial fits are given in the text. In both plots the vertical dotted line indicates the physical value of $m_{u/d}$.} 
\label{fig:fKovfpi-fK}
\end{figure*}

We work in a similar way for the determination of the ratio $f_{K}/f_{\pi}$ for which we make use of the following two fit ans\"atze: 
$f_{s\ell}/f_{\ell \ell}  = P_{0}' \, ( 1 + (5/4)\xi_{\ell} \log \xi_{\ell} + P_{1}' \xi_{\ell} + 
P_{2}' a^2 ) \,  \text{K}_{f_{K}/f_{\pi}}^{\text{FSE}}$ and  $f_{s\ell}/f_{\ell \ell}  = Q_{0}' \, \left( 1 + Q_{1}' m_{\ell} + Q_{2}' m_{\ell}^2 + Q_{3}' a^2 \right) \,  \text{K}_{f_{K}/f_{\pi}}^{\text{FSE}}$. 
In Fig.~\ref{fig:fKovfpi-fK} we show a representative plot out of  several analyses concerning the simultaneous chiral and continuum fits for $f_K$ (left panel) and $f_{K}/f_{\pi}$ (right panel). In both cases the  (NLO) SU(2) ChPT and polynomial fits to our data are of  good quality and the corresponding continuum results at the physical point are compatible. After averaging over results from all available analyses our {\it preliminary} results and the respective error budgets for the two quantities are 
\begin{align}
f_{K}^{\text{isoQCD}} & = 155.3 (0.9)_{\text{(stat+fit)}} (0.1)_{Z_P} (0.2)_{\text{chiral}} 
 (1.4)_{\text{discr.}} (0.2)_{\text{FSE}}~ [1.7]~ \text{MeV}  \label{Eq:fKresult}\\
 (f_{K}/f_{\pi})^{\text{isoQCD}} & = 1.2023 (38)_{\text{(stat+fit)}} (3)_{Z_P} (11)_{\text{chiral}} 
 (8)_{\text{discr.}} (5)_{\text{FSE}}~ [41], 
 \end{align}
where separate errors in parentheses are due to the indicated sources of uncertainty while the total error is shown in brackets. Notice that the estimate for $(f_{K}/f_{\pi})^{\text{isoQCD}}$ has a total uncertainty of about 0.34\% and it shows nice agreement with the result of Ref.~\cite{Alexandrou:2021bfr}, namely $(f_{K}/f_{\pi})^{\text{isoQCD}} = 1.1995~ (44)$, where an analysis of the same data has been performed but in terms of the PS-meson masses. Note also that a much more precise estimate for the $K$-decay constant, still in good agreement with the result of Eq.~(\ref{Eq:fKresult}), is obtained by  
$$f_{K}^{\text{isoQCD}} = (f_{K}/f_{\pi})^{\text{isoQCD}} \times f_{\pi}^{\text{(isoQCD)}} = 156.8(0.6)~\text{MeV}.$$ 
\noindent Finally, by employing the estimate  of the  strong isospin effects correction computed  in Ref.~\cite{DiCarlo:2019thl} (in the GRS scheme~\cite{GRS}) we also obtain:
$$ f_{{K}^{\pm}}/f_{{\pi}^{\pm}}  = 1.1984~ (41) ~~ \text{and}~~ f_{{K}^{\pm}}  = (f_{{K}^{\pm}}/f_{{\pi}^{\pm}}) \times f_{\pi}^{\text{(phys.)}}  = 156.3(0.6)~\text{MeV} $$

\section{Determinations of  $f_{D_{s}}$, $f_{D}$ and $f_{D_{s}}/f_{D}$  }
\begin{figure*}[h!]
\centering
\begin{tabular}{lcr}
    \includegraphics[width=0.47\linewidth]{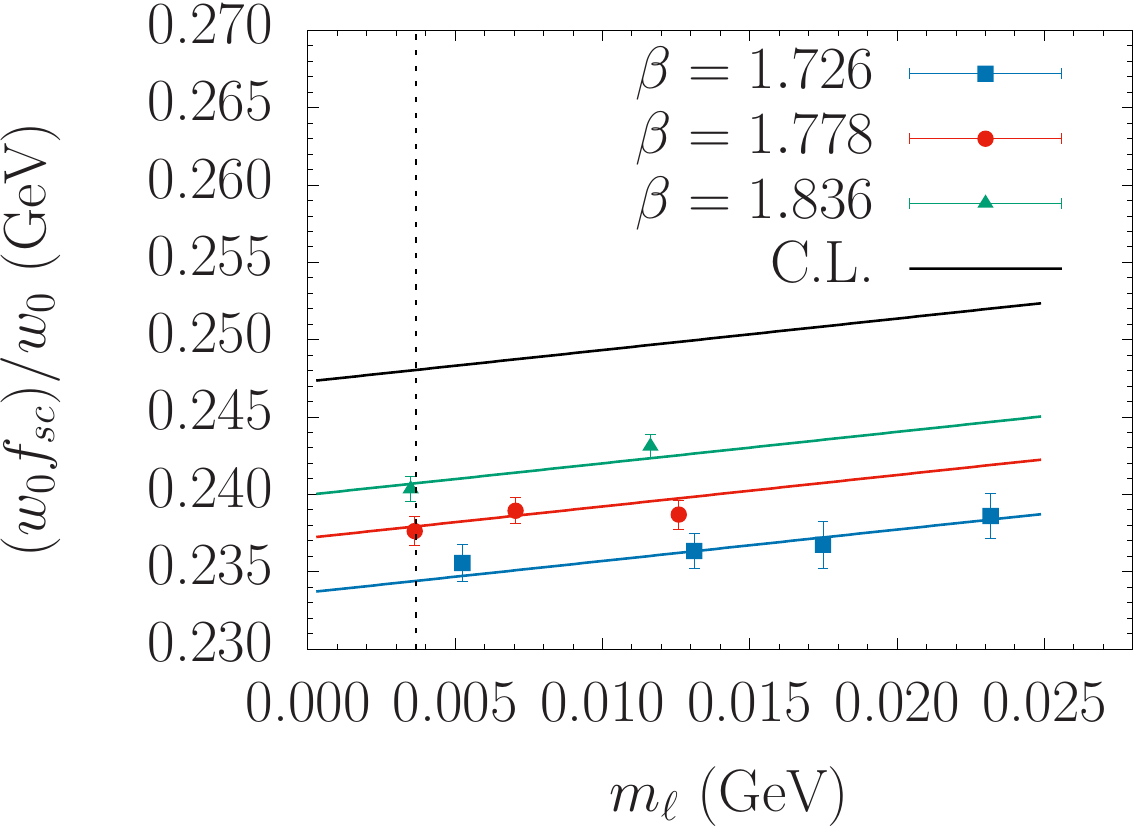}& & 
    \includegraphics[width=0.47\linewidth]{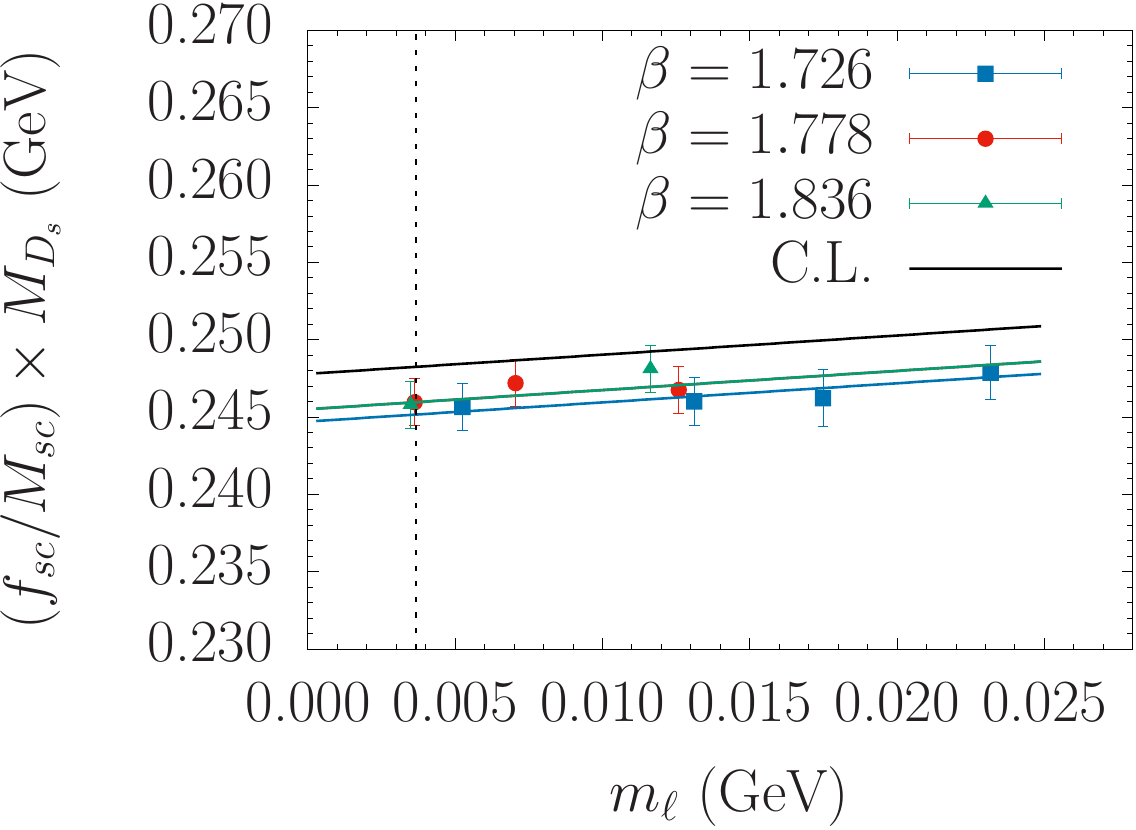} \\
    \end{tabular}
\caption{Plots showing simultaneous chiral  and continuum fit for $f_{D_{s}}$ in terms of intermediate scaling variable $w_0$  (left) and  of a PS-meson mass $M_{sc}$ (right) against the renormalised light quark mass $m_{\ell}(\overline{\text{MS}}, 2~\text{GeV})$. In both plots the vertical dotted line indicates the physical value of $m_{u/d}$ and  ``C.L.'' is for the continuum limit curve.} 
\label{fig:fDs}
\end{figure*}

We compute $f_{D_{s}}$ by first interpolating the decay constant estimates to the strange and charm quark masses before we employ a combined chiral and continuum fit of the data for  $f_{sc}$ in terms of $m_{\ell}$ and $a^2$. We try two kinds of intermediate scaling variables that are the gradient flow $w_0$ and the  pseudocalar mass $M_{sc}$, the latter computed at each value of $m_{\ell}$. The fit ansatz for both scaling variable choices is of  the form  $f_{sc} = c_{0} \, \left( 1 + c_{1} m_{\ell}  + c_3 a^2 \right)$  
that is linear in $m_{\ell}$ and it describes nicely our data as it can be appreciated by the two plots in Fig.~\ref{fig:fDs}. Dependence on $m_{\ell}$, according to the expectations, is quite weak. It should  be also added that the continuum limit results from both ways of analysis are in perfect agreement, nevertheless when  $M_{sc}$ is employed as the scaling variable cut off effects are clearly suppressed.  

\begin{figure*}[h!]
\centering
\begin{tabular}{c}
    \includegraphics[width=0.55\linewidth]{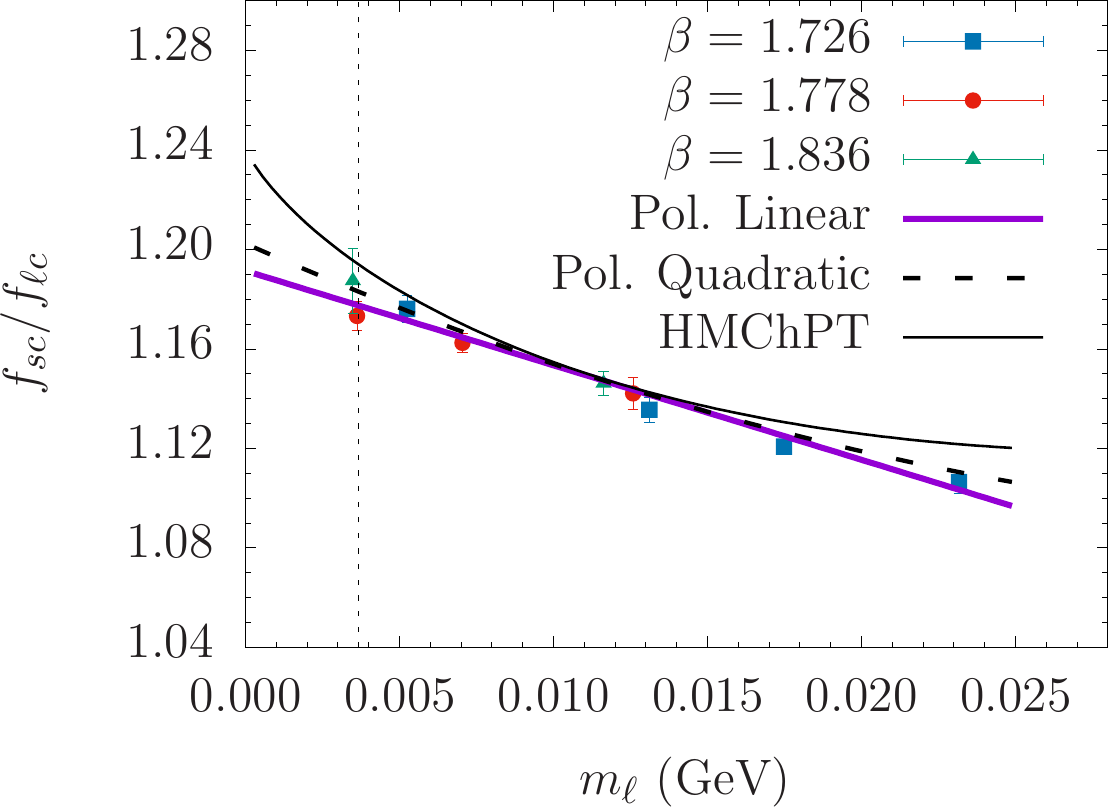} \\
    \end{tabular}
\caption{Simultaneous continuum and chiral fits (continuum limit curves are shown) for the ratio $f_{D_{s}}/f_D$ against the renormalised light quark mass $m_{\ell}(\overline{\text{MS}}, 2~\text{GeV})$ using various chiral fit ans\"atze. Vertical dotted line indicates the physical value of $m_{u/d}$.} 
\label{fig:fDsovfD}
\end{figure*}

For the calculation of the ratio $f_{D_{s}}/f_D$ we employ the following three fit ans\"atze, where the first two ans\"atze are  polynomial fits (linear or quadratic) in $m_{\ell}$, namely $f_{sc}/f_{\ell c}  = \tilde{Q}_{0} ( 1 +  \tilde Q_{1} m_{\ell} + [\tilde Q_{2} m_{\ell}^2] + \tilde Q_{3} a^2 )$,   and the third one is based on the HMChPT prediction and takes the form $f_{sc}/f_{\ell c}  =  \tilde P_{0} ( 1+  \frac{3}{4}(1+3 \hat{g}^2)\xi_{\ell} \log(\xi_{\ell}) + 
\tilde P_{1} m_{\ell} + \tilde P_{2} a^2 )$, where $\hat{g}=0.61(7)$ is obtained  from
the experimental measurement of the $g_{D^{\star} D\pi}$.  In Fig.~\ref{fig:fDsovfD} we present a representative analysis plot for the ratio $f_{D{s}}/f_{D}$ against $m_{\ell}$ where all three kinds of combined continnum and chiral fits are shown. We find that the HMCHPT fit describes rather poorly our data. Therefore having  data at (or close to) the physical point we trust as for the central value and the error only fits of good quality which in the present case are the two kinds of the  polynomial chiral fit ansatz.   
Finally,  we can compute $f_D$ directly and indirectly. In the first way, the direct one,  we follow an analysis similar to the $f_{D{s}}$ case i.e. we employ two kinds of intermediate scaling variable. As for the chiral fit ansatz we employ both polynomial and HMChPT fits. We observe that, similarly to the $f_{D_{s}}$ case, the use of a PS-meson mass of the type $M_{\ell c}$ playing the role of intermediate scaling variable leads to suppressed discretisation effects. Moreover as it happens for the analysis of the ratio $f_{D{s}}/f_{D}$, also in the case of $f_D$ the HMChPT fit ansatz does not provide a satisfactory fit quality. The indirect way for the computation of $f_D$ consists in combining our results for the ratio and the $D_{s}$ decay constants as follows $f_{D} = f_{D_{s}} / (f_{D_{s}} / f_D)$. 

Our {\it preliminary} results and error budget  read:
\begin{align}
 f_{D_{s}} & = 248.9~(1.6)_{\text{(stat+fit)}} (0.5)_{Z_P} (0.2)_{\text{chiral}} 
 (1.0)_{\text{discr.}}~[2.0]~ \text{MeV} \label{Eq:fDs_result} \\
f_D & = 210.1~(2.2)_{\text{(stat+fit)}} (0.1)_{Z_P} (0.4)_{\text{chiral}} 
  (0.8)_{\text{discr.}}~ [2.4]~\text{MeV}  \label{Eq:fD_result} \\
 f_{D_{s}}/f_{D} & = 1.1838~ (90)_{\text{(stat+fit)}} (25)_{Z_P} (38)_{\text{chiral}} 
  (57)_{\text{discr.}}~ [115],
\end{align}
where the total error for each quantity is shown in brackets. 
Notice that the total relative errors for $f_{Ds}$, $f_{D}$ and $f_{D_{s}}/f_{D}$ are 0.8\%,  1.1\% and 1.0\%, respectively.

\section{Summary and results comparisons}
In Table~\ref{Tab:Results} we provide the comparison  of the present work results (ETMC 21) with older ETMC results (ETMC 14~\cite{Carrasco:2014poa}), the latter obtained with $N_f= 2+1+1$ simulations but far from the physical point,  and also with the FLAG 19 averages~\cite{FlavourLatticeAveragingGroup:2019iem}. We would like to stress the higher precision of the ETMC 21 results with respect to the corresponding ETMC 14 ones owing to a remarkable reduction of both statistical and systematic uncertainties. We also notice that the ETMC 21 results compare well with the corresponding FLAG 19 averages.

\begin{table}
\centering
\scalebox{0.80}{
\begin{tabular}{|l|c|c|c|c|}
\hline
          &&&& \\
Quantity  & {\bf ETMC~21}   & {\bf ETMC~14}   &  {\bf FLAG~19}  & {\bf FLAG~19} \\ 
          &   ($N_f=2+1+1$)        &    ($N_f =$ 2+1+1)       &   ($N_f =$ 2+1+1) & ($N_f =$  2+1)  \\ \hline \hline

$(f_{K}/f_{\pi})^{\text{isoQCD}}$  &   1.2023(41)   &    1.188(15)     &    -  &   \\ \hline  
$f_{{K}^{\pm}}/f_{{\pi}^{\pm}}$  &   1.1984(41)   &    1.184(16)     &  1.1932(19)    &  1.1917(37) \\ \hline \hline 

$f_{K}^{\text{isoQCD}}$ (MeV) &  155.3(1.7)    &   155.0(1.9)      &   -    &  -  \\ \hline 

$f_{K}^{\text{isoQCD}} = (f_{K}/f_{\pi})^{\text{isoQCD}} \times f_{\pi}^{\text{(isoQCD)}}$ (MeV) &  156.8(0.6)    &   154.9(1.9)      &  -
& -   \\ \hline

$f_{{K}^{\pm}} = (f_{{K}^{\pm}}/f_{{\pi}^{\pm}}) \times f_{\pi}^{\text{(phys.)}}$ (MeV) &  156.3(0.6)    &   154.4(2.0)      &  155.7(0.3)
& 155.7(0.7)   \\ \hline 

$f_{D_{s}}$ (MeV) &  248.9(2.0)    &   247.2(4.1)      &  249.9(0.5)    & 248.0(1.6)  \\ \hline 

$f_{D_{s}}/f_{D}$  &   1.1838(115)   &  1.192(22)       & 1.1783(16)     & 1.1740(70)  \\ \hline  

$f_{D}$ (MeV) &  210.1(2.4)    &   207.4(3.8)      &  212.0(0.7)    &  209.0(2.4)  \\ \hline \hline 

$(\dfrac{f_{D_{s}}}{f_{D}})/(\dfrac{f_{K}}{f_{\pi}})$  &  0.995(13)    &  1.003(14)       &  -    &  -  \\  

 \hline \hline

\end{tabular}
}
\caption{Comparison of ({\it preliminary})  results for the PS-meson decay constants of the present work (ETMC 21) with previous 
ETMC results (ETMC 14~\cite{Carrasco:2014poa}) and FLAG 19 averages~\cite{FlavourLatticeAveragingGroup:2019iem}.}
\label{Tab:Results}
\end{table}

Finally, by combining  our results for the decay constants with the relevant experimental inputs  we  provide estimates for several of the first and second row elements of the CKM matrix. We get the following {\it preliminary} results: $|V_{us}/V_{ud}| = 0.2303(8)_{\text{th}} (3)_{\text{expt}} [8]$, $|V_{us}| = 0.2242(8)_{\text{th}} (3)_{\text{expt}} [8]$, $|V_{cd}| = 0.2199(25)_{\text{th}} (57)_{\text{expt}} [62]$ and $|V_{cs}| = 0.9871(79)_{\text{th}} (185)_{\text{expt}} [201]$. Thanks to the above estimates  for the unitarity checks of the first and the second CKM rows we get:
\begin{align}
 |V_{ud}|^2 + |V_{us}|^2 + |V_{ub}|^2 - 1 & = -1.56(0.34)_{\text{th}} (0.62)_{\text{expt}} [0.71] \times 10^{-3} \\ 
 |V_{cd}|^2 + |V_{cs}|^2 + |V_{cb}|^2 - 1 & = +2.3(1.6)_{\text{th}} (3.7)_{\text{expt}} [4.0] \times 10^{-2} .
\end{align}
where $|V_{ub}|^2$ and $|V_{cb}|^2$ being of order $10^{-6}$ and $10^{-4}$, respectively, have  negligible impact to the present accuracy. Our results lead to about 2$\sigma$ tension for the unitarity check of the first row  (at the per mille level)  while they confirm the second row unitarity of the CKM matrix at the percent level.

\acknowledgments
We acknowledge PRACE (Partnership for Advanced Computing in Europe) for awarding us access to 
the high-performance computing system Marconi and Marconi100 at CINECA (Consorzio Interuniversitario 
per il Calcolo Automatico dell’Italia Nord-orientale) under the grants Pra17-4394, Pra20-5171 and 
Pra22-5171, and CINECA for providing us CPU time under the specific initiative INFN-LQCD123. P.D. acknowledges support form the European Unions Horizon 2020 research and innovation programme under the Marie Sklodowska-Curie grant agreement No. 813942 (EuroPLEx) and  from INFN under the research project INFN-QCDLAT.  R.F. acknowledges the University of Rome Tor Vergata for the support granted to the project PLNUGAMMA.

\bibliography{myrefs}

\providecommand{\href}[2]{#2}\begingroup\raggedright\begin{thebibliography}{10}

\bibitem{FrezzoRoss1}
R.~Frezzotti and G.C.~Rossi, \emph{Chirally improving wilson fermions. i: O(a)
  improvement}, {\emph{JHEP} {\bfseries 08} (2004) 007}
  [\href{https://arxiv.org/abs/hep-lat/0306014}{{\ttfamily hep-lat/0306014}}].

\bibitem{Frezzotti:2003xj}
R.~Frezzotti and G.C.~Rossi, \emph{{Twisted-mass lattice QCD with mass
  non-degenerate quarks}},
  \href{https://doi.org/10.1016/S0920-5632(03)02477-0}{\emph{Nucl. Phys. Proc.
  Suppl.} {\bfseries 128} (2004) 193}
  [\href{https://arxiv.org/abs/hep-lat/0311008}{{\ttfamily hep-lat/0311008}}].

\bibitem{Alexandrou:2018egz}
[ETMC] C.~Alexandrou et~al., \emph{{Simulating twisted mass fermions at physical
  light, strange and charm quark masses}},
  \href{https://doi.org/10.1103/PhysRevD.98.054518}{\emph{Phys. Rev. D}
  {\bfseries 98} (2018) 054518}
  [\href{https://arxiv.org/abs/1807.00495}{{\ttfamily 1807.00495}}].

\bibitem{ETM:2015ned}
[ETMC] A.~Abdel-Rehim et~al., \emph{{First physics results at the physical pion mass
  from $N_f=2$ Wilson twisted mass fermions at maximal twist}},
  \href{https://doi.org/10.1103/PhysRevD.95.094515}{\emph{Phys. Rev. D}
  {\bfseries 95} (2017) 094515}
  [\href{https://arxiv.org/abs/1507.05068}{{\ttfamily 1507.05068}}].

\bibitem{Dimopoulos:2009es}
[ALPHA] P.~Dimopoulos, H.~Simma and A.~Vladikas, \emph{{Quenched B(K)-parameter from
  Osterwalder-Seiler tmQCD quarks and mass-splitting discretization effects}},
  \href{https://doi.org/10.1088/1126-6708/2009/07/007}{\emph{JHEP} {\bfseries
  0907} (2009) 007} [\href{https://arxiv.org/abs/0902.1074}{{\ttfamily
  0902.1074}}].

\bibitem{Becirevic:2006ii}
D.~Becirevic et~al., \emph{Exploring twisted mass lattice qcd with the clover
  term}, {\emph{Phys. Rev.} {\bfseries D74} (2006) 034501}
  [\href{https://arxiv.org/abs/hep-lat/0605006}{{\ttfamily hep-lat/0605006}}].

\bibitem{Iwasaki:1985we}
Y.~Iwasaki, \emph{{Renormalization Group Analysis of Lattice Theories and
  Improved Lattice Action: Two-Dimensional Nonlinear O(N) Sigma Model}},
  \href{https://doi.org/10.1016/0550-3213(85)90606-6}{\emph{Nucl.Phys.}
  {\bfseries B258} (1985) 141}.

\bibitem{Aoki:1998qd}
S.~Aoki, R.~Frezzotti and P.~Weisz, \emph{{Computation of the improvement
  coefficient c(SW) to one loop with improved gluon actions}},
  \href{https://doi.org/10.1016/S0550-3213(98)00742-1}{\emph{Nucl. Phys. B}
  {\bfseries 540} (1999) 501}
  [\href{https://arxiv.org/abs/hep-lat/9808007}{{\ttfamily hep-lat/9808007}}].

\bibitem{Alexandrou:2021bfr}
[ETMC] C.~Alexandrou et~al., \emph{{Ratio of kaon and pion leptonic decay constants
  with $N_f = 2 + 1 + 1$ Wilson-clover twisted-mass fermions}},
  \href{https://arxiv.org/abs/2104.06747}{{\ttfamily 2104.06747}}.

\bibitem{JacobLAT21}
[ETMC] J.~Finkenrath et~al., \emph{{Twisted mass gauge ensembles at physical values of
  the light, strange and charm quark masses}}, {\emph{PoS(LATTICE2021)284}
  (2021) }.

\bibitem{FlavourLatticeAveragingGroup:2019iem}
[FLAG] S.~Aoki et~al., \emph{{FLAG Review 2019: Flavour Lattice Averaging Group
  (FLAG)}}, \href{https://doi.org/10.1140/epjc/s10052-019-7354-7}{\emph{Eur.
  Phys. J. C} {\bfseries 80} (2020) 113}
  [\href{https://arxiv.org/abs/1902.08191}{{\ttfamily 1902.08191}}].

\bibitem{Osterwalder:1977pc}
K.~Osterwalder and E.~Seiler, \emph{{Gauge Field Theories on the Lattice}},
  \href{https://doi.org/10.1016/0003-4916(78)90039-8}{\emph{Ann. Phys.}
  {\bfseries 110} (1978) 440}.

\bibitem{Frezzotti:2004wz}
R.~Frezzotti and G.C.~Rossi, \emph{{Chirally improving Wilson fermions. II:
  Four-quark operators}},
  \href{https://doi.org/10.1088/1126-6708/2004/10/070}{\emph{JHEP} {\bfseries
  10} (2004) 070} [\href{https://arxiv.org/abs/hep-lat/0407002}{{\ttfamily
  hep-lat/0407002}}].

\bibitem{Alexandrou:2021gqw}
[ETMC] C.~Alexandrou et~al., \emph{{Quark masses using twisted mass fermion gauge
  ensembles}},  \href{https://arxiv.org/abs/2104.13408}{{\ttfamily
  2104.13408}}.

\bibitem{BartekLAT21}
[ETMC] B.~Kostrzewa et~al., \emph{{Gradient-flow scale setting with $N_f=2+1+1$
  Wilson-clover twisted-mass fermions}}, {\emph{PoS(LATTICE2021)131} (2021) }.

\bibitem{Aoki:2016frl}
[FLAG] S.~Aoki et~al., \emph{{Review of lattice results concerning low-energy particle
  physics}}, \href{https://doi.org/10.1140/epjc/s10052-016-4509-7}{\emph{Eur.
  Phys. J. C} {\bfseries 77} (2017) 112}
  [\href{https://arxiv.org/abs/1607.00299}{{\ttfamily 1607.00299}}].

\bibitem{DiCarlo:2019thl}
M.~Di~Carlo, D.~Giusti, V.~Lubicz, G.~Martinelli, C.T.~Sachrajda, F.~Sanfilippo
  et~al., \emph{{Light-meson leptonic decay rates in lattice QCD+QED}},
  \href{https://doi.org/10.1103/PhysRevD.100.034514}{\emph{Phys. Rev. D}
  {\bfseries 100} (2019) 034514}
  [\href{https://arxiv.org/abs/1904.08731}{{\ttfamily 1904.08731}}].

\bibitem{Frezzotti:2000nk}
R.~Frezzotti, P.A.~Grassi, S.~Sint and P.~Weisz, \emph{{Lattice QCD with a
  chirally twisted mass term}}, {\emph{JHEP} {\bfseries 0108} (2001) 058}
  [\href{https://arxiv.org/abs/hep-lat/0101001}{{\ttfamily hep-lat/0101001}}].

\bibitem{MatteoLAT21}
[ETMC] M.~Di~Carlo et~al., \emph{{Renormalization constants of quark bilinear
  operators in QCD with dynamical up, down, strange and charm quarks}},
  {\emph{PoS(LATTICE2021)399}}.

\bibitem{Colangelo:2005gd}
G.~Colangelo, S.~Durr and C.~Haefeli, \emph{{Finite volume effects for meson
  masses and decay constants}},
  \href{https://doi.org/10.1016/j.nuclphysb.2005.05.015}{\emph{Nucl. Phys. B}
  {\bfseries 721} (2005) 136}
  [\href{https://arxiv.org/abs/hep-lat/0503014}{{\ttfamily hep-lat/0503014}}].

\bibitem{GRS} J. Gasser, A. Rusetsky and I. Scimemi
 \emph{{Electromagnetic corrections in hadronic processes}},
  \href{https://doi:10.1140/epjc/s2003-01383-1}{\emph{Eur. Phys. J. }
  {\bfseries C32} (2003) 97}
  [\href{https://arxiv.org/abs/hep-ph/0305260}{{\ttfamily hep-ph/0305260 [hep-ph]}}].


\bibitem{Carrasco:2014poa}
[ETMC] N.~Carrasco et~al., \emph{{Leptonic decay constants $f_{K},f_{D},$ and
  $f_{{D}_{s}}$ with $N_{f} = 2+1+1$ twisted-mass lattice QCD}},
  \href{https://doi.org/10.1103/PhysRevD.91.054507}{\emph{Phys. Rev.}
  {\bfseries D91} (2015) 054507}
  [\href{https://arxiv.org/abs/1411.7908}{{\ttfamily 1411.7908}}].

\end{thebibliography}\endgroup

\end{document}